# Spin Triplet Ground-State in the Copper Hexamer Compounds $A_2Cu_3O(SO_4)_3$ (A=Na, K)

A. Furrer[1], A. Podlesnyak[2], E. Pomjakushina[3], and V. Pomjakushin[1]

[1] Laboratory for Neutron Scattering, Paul Scherrer Institut, CH-5232 Villigen PSI, Switzerland

[2] Neutron Scattering Division, Oak Ridge National Laboratory, Oak Ridge, Tennessee 37831, USA

[3] Laboratory for Scientific Developments and Novel Materials, Paul Scherrer Institut, CH-5232 Villigen PSI, Switzerland

The compounds $A_2Cu_3O(SO_4)_3$ (A=Na, K) are characterized by copper hexamers which are weakly coupled along the b-axis to realize one-dimensional antiferromagnetic chains below $T_N \approx 3$ K, whereas the interchain interactions along the a- and c-axes are negligible. We investigated the energy-level splittings of the copper hexamers by inelastic neutron scattering below and above $T_N$. The eight lowest-lying hexamer states could be unambiguously assigned and parametrized in terms of a Heisenberg exchange Hamiltonian, providing direct experimental evidence for an S=1 triplet ground-state associated with the copper hexamers. Therefore, the compounds $A_2Cu_3O(SO_4)_3$ serve as novel cluster-based spin-1 antiferromagnets to support Haldane's conjecture that a gap appears in the excitation spectrum below $T_N$, which was verified by inelastic neutron scattering.

Progress in quantum magnetism has often been motivated by observations on naturally occurring minerals. There are numerous examples of novel phenomena first discovered in natural samples, before they found the way to the physics laboratories [1,2]. Here we focus on the discovery of the compounds $A_2Cu_3O(SO_4)_3$ (A=Na, K). The minerals fedotovite (A=K) [3] and euchlorine ($A_2$=KNa) [4] were found in sublimates of the Tolbachik fission eruption (Kamchatka peninsula, Russia) in the years 1975-1976, which was complemented at the same place in the years 2014-2015 by the mineral puninite (A=Na) [5]. All these minerals are built up of edge-shared tetrahedral spin clusters consisting of six $Cu^{2+}$ ions with S=1/2 spins as schematically shown in Fig. 1(a). The hexamer $Cu^{2+}$ clusters are magnetically decoupled along the a and c axes due to the lack of exchange paths, but they are weakly coupled along the b axis giving rise to one-dimensional antiferromagnetic order below $T_N \approx 3$ K. The fascinating properties of the compounds $A_2Cu_3O(SO_4)_3$ have only very recently been recognized by Fujihara *et al.* [6], who carried out experimental studies for A=K by magnetic susceptibility, magnetization, heat capacity, and inelastic neutron scattering (INS) measurements (restricted to energy transfers below 2.5 meV). The presence of a spin triplet ground-state was suggested, which puts fedotovite into a novel Haldane state based on spin-cluster chains. Haldane's conjecture [7] predicts gapless excitations for half-integer spin systems, whereas a gap opens for integer spin systems.

The postulated spin S=1 ground-state in $K_2Cu_3O(SO_4)_3$ essentially relied on an analysis of thermodynamic magnetic properties [6]. However, for large complexes such as the $Cu^{2+}$ hexamers, the latter can be rationalized by a variety of different sets of magnetic exchange parameters, so that the energy-level scheme of the $Cu^{2+}$ hexamers cannot be unambiguously determined. Consequently, the spin triplet ground-ground state postulated in Ref. [6] has to be considered as a hypothesis missing the criteria of being both necessary and sufficient. More specifically, the spin states of $Cu^{2+}$ hexamers comprise five singlets (S=0), nine triplets (S=1), five quintets (S=2), and a septet (S=3) [8], but in principle none of these states can be excluded from being the true ground state. What is needed are spectroscopic methods such as the INS technique to allow a direct determination of the spin states. Therefore, we performed INS experiments for the compounds $A_2Cu_3O(SO_4)_3$ (A=Na, K) in order to arrive at a detailed picture of the spin excitations associated with the $Cu^{2+}$ hexamers, thereby confirming directly the spin triplet ground-state.

The compounds $A_2Cu_3O(SO_4)_3$ (A=Na, K) crystallize in the monoclinic space group C2/c. Polycrystalline samples were synthesized by a solid-state reaction process as described in Ref. [6]. The samples were characterized by X-ray and neutron diffraction, confirming their single-phase character [9]. Table I lists the lattice parameters and the Cu-O-Cu bond angles determined by neutron diffraction at T=2 K. The bond angles exhibit relevant asymmetries with respect to the central Cu3-Cu3 bond.

INS experiments were carried out with use of the high-resolution time-of-flight spectrometer CNCS [10] at the spallation neutron source (SNS) at Oak Ridge National Laboratory. Fig. 2 shows energy spectra observed for $A_2Cu_3O(SO_4)_3$ which exhibit similar features for both A=Na and A=K as expected from the similar structural parameters. Prominent peak features appear in three windows of energy transfers ΔE: 4<ΔE<12 meV (I), 12<ΔE<19 meV (II), and 24<ΔE<34 meV (III). No peak-like features were observed for ΔE>35 meV.

The signals in window I increase upon increasing both the temperature T (see Fig. 2) and the modulus of the scattering vector **Q**, which is characteristic of phonon scattering. In window II phonon scattering is still present, but on top of it there are three narrow peaks (denoted as $S_1$, $S_2$, $S_3$), whose intensities decrease with increasing Q according to the square of the magnetic form factor $F^2(Q)$ [9], so that we associate them with spin excitations of the $Cu^{2+}$ hexamers. This interpretation is furthermore supported by the temperature dependence.

When increasing the temperature from 1.5 K to 6 K, the three peaks are shifted downwards by typically 0.5 meV (see Fig. 2), which results from the opening of a gap below $T_N$. In window III we observe four partially resolved peaks (denoted as $T_2$, $Q_1$, $T_3$, $T_4$) whose positions are determined by Gaussian least-squares fits. The Q-dependence of the four peaks follows the square of the magnetic form factor $F^2(Q)$ [9], thereby confirming their origin in terms of spin excitations of the $Cu^{2+}$ hexamers. Upon raising the temperature, all the spin excitations observed in windows II and III gradually lose intensity according to Boltzmann statistics, so that we can associate them with ground-state transitions. In addition, we observe a considerable line broadening of the spin excitations with increasing temperature due to relaxation effects. Raising the temperature from 6 K to 20 K and 60 K results in an increase of the line widths by a factor 2 and 3, respectively, which prevents the observation of excited-state $Cu^{2+}$ hexamer transitions.

The energies of the spin excitations observed for $A_2Cu_3O(SO_4)_3$ (A=Na, K) are summarized in Fig. 1(b). We analyze these results in terms of a Heisenberg spin Hamiltonian which for $Cu^{2+}$ hexamers has the form

$$H = -2 \sum_{i,j=1}^{6} J_{ij} \mathbf{S_i} \cdot \mathbf{S_j} \quad , \tag{1}$$

where $J_{ij}$ and $\mathbf{S_i}$ denote the bilinear exchange parameters and the spin operators of the $Cu^{2+}$ ions, respectively. The parameters $J_{ij}$ originate from superexchange interactions provided by the O1 and O12 ions situated in the centers of the two Cu tetrahedra. The nature of the superexchange interaction is strongly dependent on the Cu-O-Cu bond angle [11,12]. Ferromagnetic coupling is obtained for bond angles around 90°. With increasing bond angle the ferromagnetic coupling is gradually weakened and turns into an antiferromagnetic coupling.

Eq. (1) gives rise to twenty $Cu^{2+}$ hexamer states, namely five singlets ($S_i$), nine triplets ($T_i$), five quintets ($Q_i$), and one septet. The energy level splitting is in agreement with an analysis of spin-1/2 hexamers consisting of two isosceles trimers [8], but this coupling scheme is not realized for the compounds $A_2Cu_3O(SO_4)_3$ (A=Na, K). In principle, Eq. (1) involves fifteen independent exchange parameters $J_{ij}$, whose number can be reduced to seven individual parameters $J_{ij}$ by confining to the leading interactions as well as by using the symmetry properties of the Cu-O-Cu bond angles listed in Table I. In order to avoid spin frustration within the $Cu^{2+}$ hexamers, the coupling scheme involves three ferromagnetic interactions $J_{cd} > J_{ab}, J_{cf}$ and four antiferromagnetic interactions $|J_{ad}|, |J_{cf}| > |J_{ac}|, |J_{ce}|$ as defined in Fig. 1(a). The nature as well as the relative sizes of the parameters $J_{ij}$ are dictated by the Cu-O-Cu bond angles. This coupling scheme gives rise to a triplet ground-state ($T_1$), separated from a group of three excited singlet states ($S_1$, $S_2$, $S_3$) and further separated by a group of excited triplet and quintet states ($T_2$, $Q_1$, $T_3$, $T_4$), in nice agreement with the observations (see Fig. 2).

A least-squares fitting procedure based on Eq. (1) was applied to the observed excitation energies displayed in Fig. 1(b). The resulting exchange parameters are listed in Table I. The parameter $J_{cd}$ turns out to have little influence on the observed excitation energies, which explains its large experimental uncertainty. We find good agreement between the observed and calculated energies with $\chi^2=1.1$ and $\chi^2=1.8$ for A=Na and A=K, respectively. The differences of the exchange parameters derived for A=Na and K are rather small because of the robust structure of the $Cu^{2+}$ tetrahedra. The assignments of the excited states are also supported by intensity calculations [9]. In particular, almost equal intensities are predicted for the transitions both to the three singlet states $S_1$, $S_2$, $S_3$ and to the three triplet states $T_2$, $T_3$, $T_4$ as experimentally observed. The transition to the quintet state $Q_1$ is calculated to be about twice as intense as the transitions to the triplet states $T_2$, $T_3$, $T_4$, in agreement with the observed intensities displayed in Fig. 2.

Our parameters predict the overall energy splittings of the $Cu^{2+}$ hexamers to be 67 meV (for A=Na) and 66 meV (for A=K). Our INS experiments revealed spin excitations up to energy transfers of 32 meV. Higher-lying spin excitations could not be detected due to either small transition matrix elements or transitions forbidden by the dipole selection-rules. Nevertheless, we emphasize as the most important conclusion that our analysis of the experimental data clearly results in a triplet S=1 ground state of the $Cu^{2+}$ hexamers.

The opening of a spin gap below $T_N$ was verified by INS experiments as shown in Fig. 3. At T=1.5 K there is clear evidence of gapped energy spectra for both compounds $A_2Cu_3O(SO_4)_3$ (A=Na, K). The energies of the dispersive modes associated with the S=1 ground state extend from 0.6 meV to 1.7 meV. From the corresponding width (1.1 meV) we estimate the antiferromagnetic intercluster exchange interaction provided by the superexchange path Cu1-O-S-O-Cu2 to be $2J_{inter}$=-1.1 meV, which is much smaller than the intracluster parameters $J_{ij}$, since the superexchange involves three non-magnetic ions. The lower bound of the gap (0.6 meV) is consistent with the downward shift of the spin excitations $S_1$, $S_2$, $S_3$ when going from T=1.5 K to T=6 K, see Fig. 2. At T=3 K the gap is already closed as shown in Fig. 3. The observed intensity above $T_N$ is characteristic of short-range antiferromagnetic correlations associated with the one-dimensional $Cu^{2+}$ hexamer chains. Our results are in agreement with INS data obtained for A=K by Fujihara *et al.* [6].

In conclusion, our analysis of INS spectroscopy data provided direct experimental evidence for the existence of spin triplet ground-states in the $Cu^{2+}$ hexamer compounds $A_2Cu_3O(SO_4)_3$ (A=Na, K). A reliable set of seven parameters describing the $Cu^{2+}$ exchange interactions was derived which considerably differs from the approach adopted in Ref. [6]. The case of edge-shared tetrahedral spin clusters has been generalized in the sense that the ground-state alternates between singlet and triplet depending on the number of tetrahedra, *i.e.,* any cluster compound with an even number of tetrahedra would realize a Haldane state [6,7]. This rule is nicely confirmed when we apply the exchange parameters obtained in the present work to the case of a single $Cu^{2+}$ tetrahedron.

Finally we mention that nature is rich in minerals containing oxocentric $Cu^{2+}$ tetrahedra ($OCu_4$) as basic constituents in a variety of configurations, ranging from three-dimensional crystal frameworks and two-dimensional layer systems to one-dimensional chains and isle-like complexes discussed in the present work. The topic of one-dimensional chains is of particular significance due to the competition between intra- and inter-cluster exchange interactions, which opens interesting perspectives for spin-frustrated systems such as the emergence of spinon excitations in spin chains [13]. We mention as examples the compounds $Cu_2OCl_2$ [14], $Cu_3Mo_2O_9$ [15], $AgCuVO_4$ [16], $K_3Cu_3AlO_2(SO_4)_4$ [17], and $KCu_3OCl(SO_4)_2$ [18], which are characterized by chains of corner-sharing tetrahedra, as well as $Li_2ZrCuO_4$ [19] formed by chains of edge-shared tetrahedra. We are not aware of any experimental work describing the magnetic interactions within a $OCu_4$ tetrahedron in detail. Since the $OCu_4$ constituents are structurally rather stable units, the generic exchange coupling scheme determined in the present work for $A_2Cu_3O(SO_4)_3$ (A=Na, K) will be a reliable basis for application to the large number of related compounds.

Discussions with H. U. Güdel (University of Berne, Switzerland) are gratefully acknowledged. Part of this work was performed at the Swiss Spallation Neutron Source (SINQ), Paul Scherrer Institut (PSI), Villigen, Switzerland. This research used resources at the Spallation Neutron Source, a

TABLE I. Lattice parameters (a,b,c,β), Cu-O-Cu bond angles (Θ), and exchange parameters ($J_{ij}$ as defined in Fig. 1a) of the compounds $A_2Cu_3O(SO_4)_3$ (A=Na, K) determined at T=2 K.

| | $Na_2Cu_3O(SO_4)_3$ | $K_2Cu_3O(SO_4)_3$ |
|---|---|---|
| a [Å] | 17.21406(67) | 18.97550(66) |
| b [Å] | 9.37286(35) | 9.50038(35) |
| c [Å] | 14.37014(54) | 14.19721(51) |
| β [°] | 111.84364(75) | 110.49150(85) |
| Θ($Cu2_a$-O12-$Cu2_b$) [°] | 105.2(4) | 102.2(5) |
| Θ($Cu2_a$-O12-$Cu3_{c/d}$) [°] | 106.5(4) / 124.4(4) | 107.0(5) / 124.2(5) |
| Θ($Cu2_b$-O12-$Cu3_{d/c}$) [°] | 106.5(4) / 124.4(4) | 107.0(5) / 124.2(5) |
| Θ($Cu3_c$-O1/O12-$Cu3_d$) [°] | 92.7(4) / 91.2(4) | 93.0(5) / 94.0(5) |
| Θ($Cu3_c$-O1-$Cu1_{e/f}$) [°] | 110.6(4) / 121.3(5) | 111.7(5) / 123.0(6) |
| Θ($Cu3_d$-O1-$Cu1_{f/e}$) [°] | 110.6(4) / 121.3(5) | 111.7(5) / 123.0(6) |
| Θ($Cu1_e$-O1-$Cu1_f$) [°] | 101.7(4) | 96.8(5) |
| $J_{ab}$ [meV] | 1.6(3) | 1.7(4) |
| $J_{ac}$ [meV] | -3.5(3) | -3.7(4) |
| $J_{ad}$ [meV] | -9.9(2) | -9.5(3) |
| $J_{cd}$ [meV] | 5(2) | 4(2) |
| $J_{ce}$ [meV] | -4.7(3) | -4.0(3) |
| $J_{cf}$ [meV] | -8.1(4) | -8.5(3) |
| $J_{ef}$ [meV] | 2.0(3) | 1.9(3) |

**FIGURE CAPTIONS**

FIG. 1. (Color online) (a): Schematic picture of the $Cu^{2+}$ hexamers in the compounds $A_2Cu_3O(SO_4)_3$ (A=Na, K). The parameters $J_{ij}$ denote the exchange coupling scheme adopted to describe the observed spin excitations. (b): Energies of the spin excitations observed for $A_2Cu_3O(SO_4)_3$ (A=Na, K) denoted by $S_i$, $T_i$, and $Q_i$ for the singlet, triplet, and quintet states, respectively.

FIG. 2. (Color online) Energy spectra of neutrons scattered from $A_2Cu_3O(SO_4)_3$ (A=Na, K). The arrows mark the transitions associated with the excited $Cu^{2+}$ hexamer states. The lines in the right-hand-side panels correspond to Gaussian least-squares fits.

FIG. 3. (color online) Energy spectra of neutrons scattered from $A_2Cu_3O(SO_4)_3$ (A=Na, K). The dashed curves correspond to the tail of the elastic line.

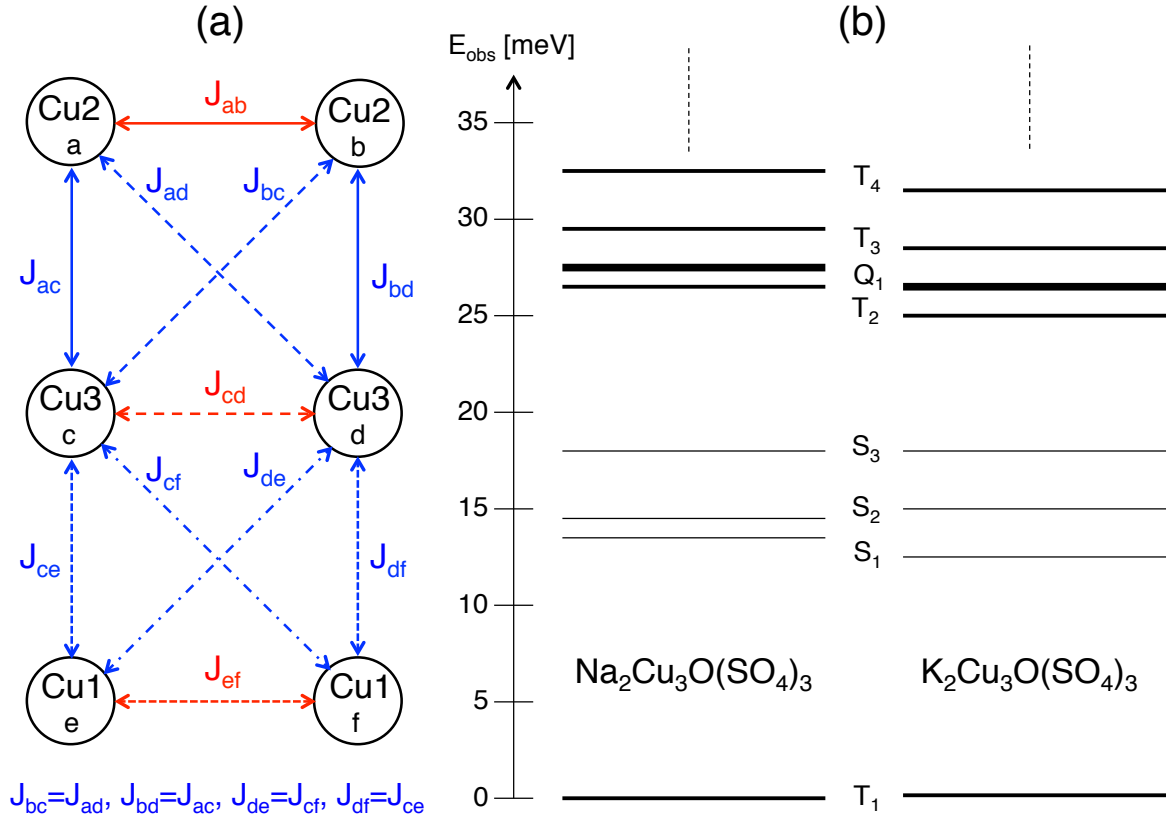

FIG. 1. (Color online) (a): Schematic picture of the $Cu^{2+}$ hexamers in the compounds $A_2Cu_3O(SO_4)_3$ (A=Na, K). The parameters $J_{ij}$ denote the exchange coupling scheme adopted to describe the observed spin excitations. (b): Energies of the spin excitations observed for $A_2Cu_3O(SO_4)_3$ (A=Na, K) denoted by $S_i$, $T_i$, and $Q_i$ for the singlet, triplet, and quintet states, respectively.

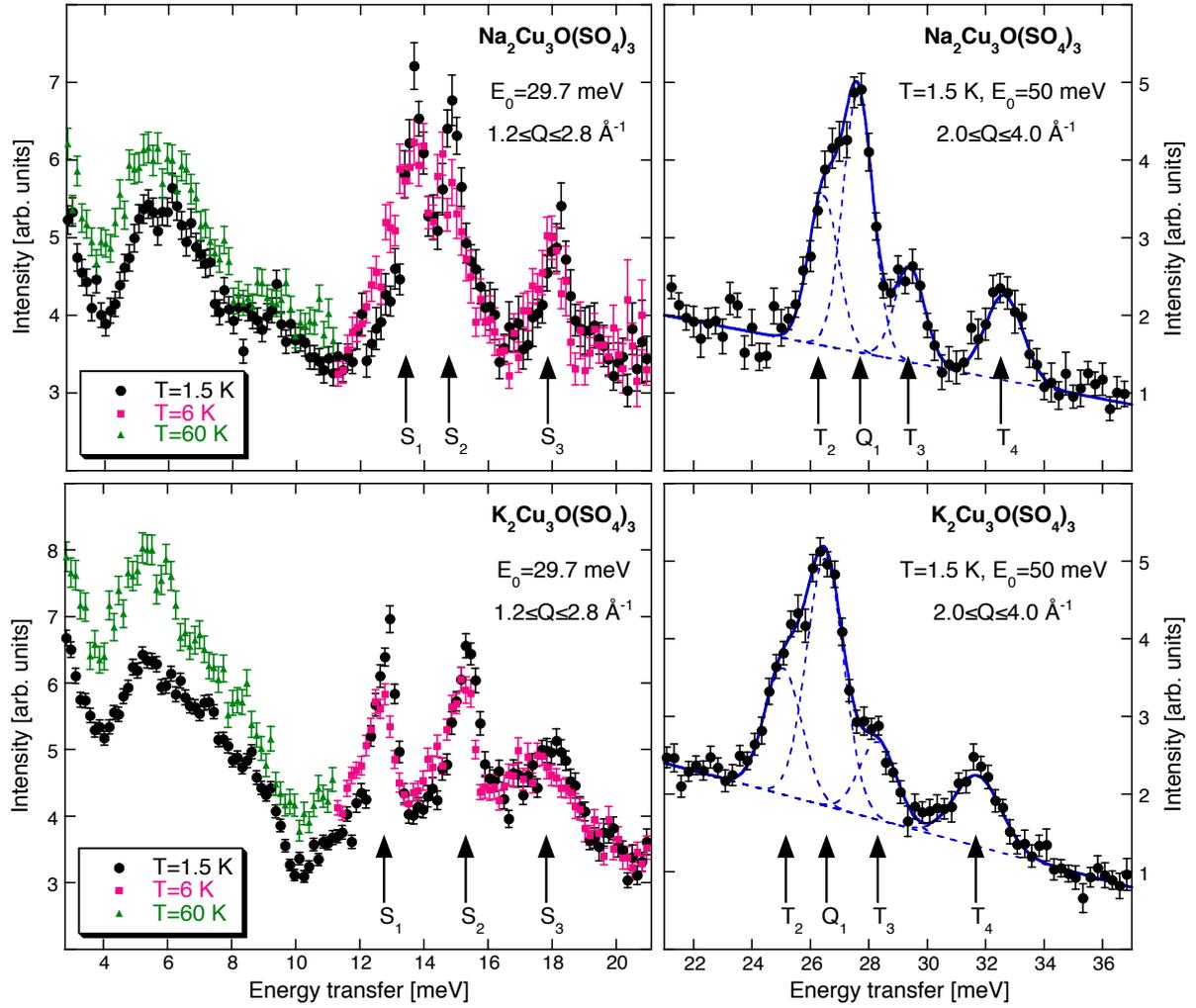

FIG. 2. (Color online) Energy spectra of neutrons scattered from $A_2Cu_3O(SO_4)_3$ (A=Na, K). The arrows mark the transitions associated with the excited $Cu^{2+}$ hexamer states. The lines in the right-hand-side panels correspond to Gaussian least-squares fits.

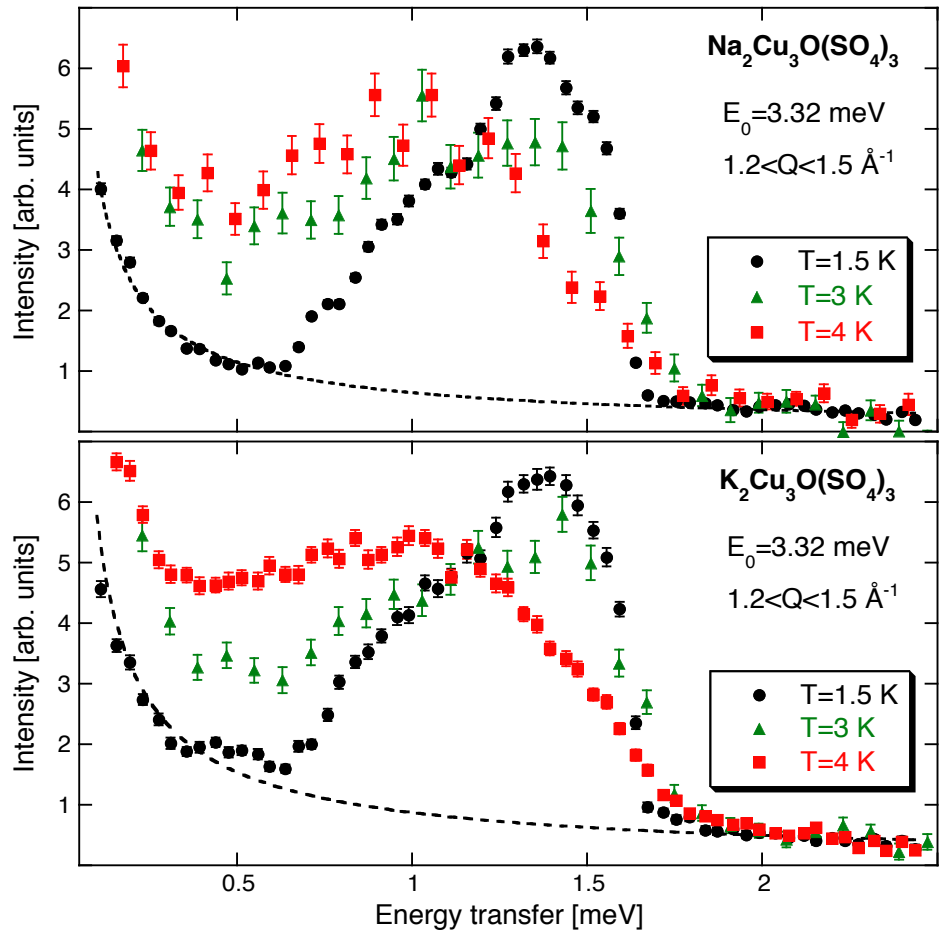

FIG. 3. (color online) Energy spectra of neutrons scattered from $A_2Cu_3O(SO_4)_3$ (A=Na, K). The dashed curves correspond to the tail of the elastic line.

# Supplemental Material for
# Spin Triplet Ground-State in the Copper Hexamer Compounds $A_2Cu_3O(SO_4)_3$ (A=Na, K)

A. Furrer[1], A. Podlesnyak[2], E. Pomjakushina[3], and V. Pomjakushin[1]

[1] Laboratory for Neutron Scattering, Paul Scherrer Institut, CH-5232 Villigen PSI, Switzerland

[2] Neutron Scattering Division, Oak Ridge National Laboratory, Oak Ridge, Tennessee 37831, USA

[3] Laboratory for Scientific Developments and Novel Materials, Paul Scherrer Institut, CH-5232 Villigen PSI, Switzerland

## 1. EXPERIMENTAL DETAILS

**1.1. Sample synthesis and characterization**

The compounds $A_2Cu_3O(SO_4)_3$ (A=Na, K) crystallize in the monoclinic space group C2/c. Polycrystalline samples were synthesized by a solid-state reaction process. High-purity CuO, $CuSO_4$, and $A_2SO_4$ (A=Na, K) were mixed in a molar ratio 1:2:1, followed by annealing at 500ºC and 580ºC in air (for Na and K, respectively) for at least 100 h with intermediate grindings. X-ray diffraction confirmed the single-phase character of the samples. Magnetic susceptibility measurements were performed with use of a Quantum Design Magnetic Properties Measurement System for temperatures 2<T<300 K. The inverse susceptibilities exhibit a kink around T=120 K for both A=Na and K, similar to the data obtained for A=K [1].

## 1.2. Neutron Diffraction

The neutron powder diffraction experiments were performed with use of the high-resolution diffractometer for thermal neutrons, HRPT [2], at the spallation neutron source SINQ at PSI Villigen with neutron wavelength λ=2.45 Å at temperatures T=2 K, 150 K, and 300 K. The refinements of the crystal structures were carried out with the program FULLPROF [3]. The results obtained at T=2 K are listed in Table S1. The lattice parameters at T=150 K and 300 K are as follows:

$Na_2Cu_3O(SO_4)_3$:

T=150 K: a=17.24669(60) Å, b= 9.38066(32) Å, c=14.37664(50) Å, β=111.89604(75)°

T=300 K: a=17.31714(48) Å, b= 9.39965(25) Å, c=14.39339(39) Å, β=111.95650(60)°

$K_2Cu_3O(SO_4)_3$:

T=150 K: a=19.01577(57) Å, b= 9.50959(31) Å, c=14.19340(45) Å, β=110.55505(86)°

T=300 K: a=19.08335(58) Å, b= 9.52801(32) Å, c=14.20051(44) Å, β=110.60856(82)°

## 1.3. Inelastic Neutron Scattering

The inelastic neutron scattering (INS) experiments were carried out with use of the high-resolution time-of-flight spectrometer CNCS [4] at the spallation neutron source (SNS) at Oak Ridge National Laboratory. The samples were enclosed in aluminum cylinders (8 mm diameter, 45 mm height) and placed into a He cryostat to achieve temperatures T≥1.5 K. Additional experiments were performed for vanadium to allow the correction of the raw data with respect to background, detector efficiency, and absorption according to standard procedures. The measurements were performed for incoming neutron energies $E_0$=3.32, 29.7, and 50 meV and temperatures 1.5≤T≤60 K.

## 2. NEUTRON CROSS-SECTION

The neutron cross-section for $Cu^{2+}$ hexamer transitions can be derived from the general formula for magnetic neutron scattering [5]:

$$\frac{d^2\sigma}{d\Omega d\omega} \propto F^2(\mathbf{Q}) \sum_{\alpha,\beta} \left( \delta_{\alpha\beta} - \frac{Q_\alpha Q_\beta}{Q^2} \right) S^{\alpha\beta}(\mathbf{Q},\omega)$$

with (1)

$$S^{\alpha\beta}(\mathbf{Q},\omega) = \sum_{i,j=1}^{6} \exp\{i\mathbf{Q} \cdot (\mathbf{R_i} - \mathbf{R_j})\} \sum_{\lambda,\lambda'} p_\lambda \langle \lambda | S_i^\alpha | \lambda' \rangle \langle \lambda' | S_j^\beta | \lambda \rangle \ \delta(\hbar\omega + E_\lambda - E_{\lambda'})$$

where F(**Q**) is the magnetic form factor, **Q** the scattering vector, and $S_i^\alpha$ (α=x,y,z) the spin operator of the i*th* ion at site $\mathbf{R_i}$. |λ> denotes the initial state of the system, with energy $E_\lambda$ and thermal population factor $p_\lambda$, and its final state is |λ'>. <λ|$S_i^\alpha$|λ'> and <λ'|$S_j^\beta$|λ> are the transition matrix elements. For $Cu^{2+}$ hexamers with spin $S_i$=1/2, the state |λ> is defined by a combination of all possible values of the secondary spin quantum numbers $M_i$ with -$S_i$≤$M_i$≤$S_i$, giving rise to $(2S_i+1)^6$=64 components. Each hexamer state |λ> is characterized by the total spin quantum number $M_\lambda$. We have $M_\lambda$=0 for singlets, -1≤$M_\lambda$≤1 for triplets, -2≤$M_\lambda$≤2 for quintets, and -3≤$M_\lambda$≤3 for septets. The transitions |λ>→|λ'> are governed by the dipole selection rules ΔM=$M_\lambda$-$M_{\lambda'}$=0 and ΔM=$M_\lambda$-$M_{\lambda'}$=±1.

For polycrystalline material Eq. (1) has to be averaged in **Q** space, which concerns the polarization factor ($\delta_{\alpha\beta}$-$Q_\alpha Q_\beta$/$Q^2$) and the structure factor

exp{i**Q**·(**R**$_i$-**R**$_j$)}. The Q-averaging procedure results in damped oscillatory Q-dependences of the intensities for ΔM=0 and ΔM=±1 transitions [6]:

$$\left\langle \frac{d^2\sigma}{d\Omega d\omega} \right\rangle^{\Delta M=0}_{Q-average} \propto F^2(Q) \sum_{i,j=1}^{6} \left[ \frac{2}{3} + \frac{2\cos(QR_{ij})}{Q^2 R_{ij}^2} - \frac{2\sin(QR_{ij})}{Q^3 R_{ij}^3} \right] \sum_{\lambda,\lambda'} p_\lambda T_{\lambda\lambda'}$$

(2)

$$\left\langle \frac{d^2\sigma}{d\Omega d\omega} \right\rangle^{\Delta M=\pm 1}_{Q-average} \propto F^2(Q) \sum_{i,j=1}^{6} \left[ \frac{2}{3} - \frac{\cos(QR_{ij})}{Q^2 R_{ij}^2} + \frac{\sin(QR_{ij})}{Q^3 R_{ij}^3} - \frac{\sin(QR_{ij})}{QR_{ij}} \right] \sum_{\lambda,\lambda'} p_\lambda T_{\lambda\lambda'}$$

with $R_{ij}$=|**R**$_i$-**R**$_j$|. $T_{\lambda\lambda'}$ stands for the transition matrix elements defined in Eq. (1). Fig. S1 shows the Q-averaged cross-sections by setting all the matrix elements $T_{\lambda\lambda'}$ equal to 1. The ΔM=0 and ΔM=±1 cross sections have maxima around 1.1 Å$^{-1}$ and 1.4 Å$^{-1}$, respectively, and they decrease with larger Q-values roughly in accordance with $F^2(Q)$. For Q>1.5 Å$^{-1}$ the damped oscillatory features given by Eq. (2) are more or less smeared out due to the large number of 36 individual $S^{\alpha\beta}(\mathbf{Q},\omega)$ terms in Eq. (1). When we use the matrix elements associated with a particular transition |λ>→|λ'>, the overall shape of the cross sections displayed in Fig. S1 is roughly maintained, but the intensities are scaled according to the effective sizes of the matrix elements. The oscillatory Q-dependence of the cross section usually serves as a strong criterion for the assignment of magnetic cluster excitations. In the present experiments, however, the low-Q range (Q<1.5 Å$^{-1}$) was not accessible due to kinematical constraints imposed by the large spin excitation energies ≥13 meV.

We calculated the integrated intensities of the transitions $S_1$, $S_2$, $S_3$ and $T_2$, $Q_1$, $T_3$, $T_4$ observed for $K_2Cu_3O(SO_4)_3$ at T=1.5 K according to Eqs. (1) and (2) and scaled them to the Q-averaged cross section. As can be seen from Fig. S1, there is good agreement between the observed and the calculated intensities.

TABLE S1. Structure parameters (lattice parameters a, b, c, β; fractional atomic coordinates x, y, z; isotropic displacement factor B; reliability factors $R_n$ and $\chi^2$ defined in Ref. [3]) of the compounds $A_2Cu_3O(SO_4)_3$ (A=Na, K) determined at T=2 K in the monoclinic space group C2/c.

| | | $Na_2Cu_3O(SO_4)_3$ | $K_2Cu_3O(SO_4)_3$ |
|---|---|---|---|
| a [Å] | | 17.21406(67) | 18.97550(66) |
| b [Å] | | 9.37286(35) | 9.50038(35) |
| c [Å] | | 14.37014(54) | 14.19721(51) |
| β [°] | | 111.84364(75) | 110.49150(85) |
| Cu1 | x / y / z / B[Å$^2$] | 0.47740(24) / 0.02021(53) / 0.34063(31) / 0.416(62) | 0.48127(27) / 0.02265(54) / 0.34025(32) / 0.107(70) |
| Cu2 | x / y / z / B[Å$^2$] | 0.48559(23) / 0.47886(44) / 0.14103(31) / 0.416(62) | 0.48577(23) / 0.47775(52) / 0.13924(33) / 0.107(70) |
| Cu3 | x / y / z / B[Å$^2$] | 0.41231(25) / 0.74746(49) / 0.20169(33) / 0.416(62) | 0.41989(28) / 0.74506(65) / 0.20554(35) / 0.107(70) |
| S1 | x / y / z / B[Å$^2$] | 0.51074(69) / 0.75095(132) / 0.49084(90) / 0.100(0) | 0.51025(83) / 0.75760(137) / 0.48794(103) / 0.000(0) |
| S2 | x / y / z / B[Å$^2$] | 0.66266(68) / 0.03545(119) / 0.36881(69) / 0.100(0) | 0.64792(67) / 0.01631(144) / 0.36716(88) / 0.000(0) |
| S3 | x / y / z / B[Å$^2$] | 0.33644(69) / 0.46304(134) / 0.21219(73) / 0.100(0) | 0.35346(73) / 0.46326(129) / 0.21577(75) / 0.000(0) |
| Na/K1 | x / y / z / B[Å$^2$] | 0.32643(62) / 0.77745(115) / 0.44987(70) / 1.976(177) | 0.32833(50) / 0.75594(122) / 0.44147(74) / 0.000(0) |
| Na/K2 | x / y / z / B[Å$^2$] | 0.19742(59) / 0.75664(114) / 0.14412(73) / 1.976(177) | 0.19808(54) / 0.74326(107) / 0.12717(65) / 0.000(0) |
| O1 | x / y / z / B[Å$^2$] | 0.50000(0) / 0.89069(88) / 0.25000(0) / 0.470(48) | 0.50000(0) / 0.88763(110) / 0.25000(0) / 0.809(55) |
| O2 | x / y / z / B[Å$^2$] | 0.44621(35) / 0.82980(54) / 0.40845(43) / 0.470(48) | 0.45064(39) / 0.82620(66) / 0.40619(51) / 0.809(55) |
| O3 | x / y / z / B[Å$^2$] | 0.56979(37) / 0.67387(57) / 0.45829(42) / 0.470(48) | 0.56313(38) / 0.67920(67) / 0.45639(58) / 0.809(55) |
| O4 | x / y / z / B[Å$^2$] | 0.46056(37) / 0.64844(65) / 0.52811(45) / 0.470(48) | 0.46599(37) / 0.64458(71) / 0.53105(63) / 0.809(55) |
| O5 | x / y / z / B[Å$^2$] | 0.59726(38) / 0.05441(52) / 0.41646(47) / 0.470(48) | 0.58986(33) / 0.05567(66) / 0.41090(49) / 0.809(55) |
| O6 | x / y / z / B[Å$^2$] | 0.39434(36) / 0.44072(57) / 0.31844(42) / 0.470(48) | 0.40609(32) / 0.44216(69) / 0.32325(44) / 0.809(55) |
| O7 | x / y / z / B[Å$^2$] | 0.32004(32) / 0.61493(74) / 0.18793(40) / 0.470(48) | 0.33779(39) / 0.61890(77) / 0.20313(46) / 0.809(55) |
| O8 | x / y / z / B[Å$^2$] | 0.25932(39) / 0.38618(62) / 0.19212(45) / 0.470(48) | 0.28062(39) / 0.39530(63) / 0.19823(50) / 0.809(55) |
| O9 | x / y / z / B[Å$^2$] | 0.55949(34) / 0.84761(61) / 0.57348(40) / 0.470(48) | 0.54876(33) / 0.85156(70) / 0.57034(47) / 0.809(55) |
| O10 | x / y / z / B[Å$^2$] | 0.63274(33) / 0.09045(56) / 0.26494(43) / 0.470(48) | 0.62012(36) / 0.06871(67) / 0.25477(45) / 0.809(55) |
| O11 | x / y / z / B[Å$^2$] | 0.37824(34) / 0.40154(55) / 0.14432(46) / 0.470(48) | 0.38531(39) / 0.41091(66) / 0.14505(47) / 0.809(55) |
| O12 | x / y / z / B[Å$^2$] | 0.50000(0) / 0.60046(100) / 0.25000(0) / 0.470(48) | 0.50000(0) / 0.60514(118) / 0.25000(0) / 0.809(55) |
| O13 | x / y / z / B[Å$^2$] | 0.67829(34) / 0.87542(69) / 0.36813(42) / 0.470(48) | 0.65872(33) / 0.87235(76) / 0.36816(48) / 0.809(55) |
| O14 | x / y / z / B[Å$^2$] | 0.73655(42) / 0.10536(55) / 0.43271(45) / 0.470(48) | 0.71665(35) / 0.09730(70) / 0.42458(48) / 0.809(55) |
| $R_p$[%] / $R_{wp}$[%] / $R_{exp}$[%] / $\chi^2$ | | 3.39 / 4.45 / 2.99 / 2.20 | 3.58 / 4.65 / 3.00 / 2.41 |

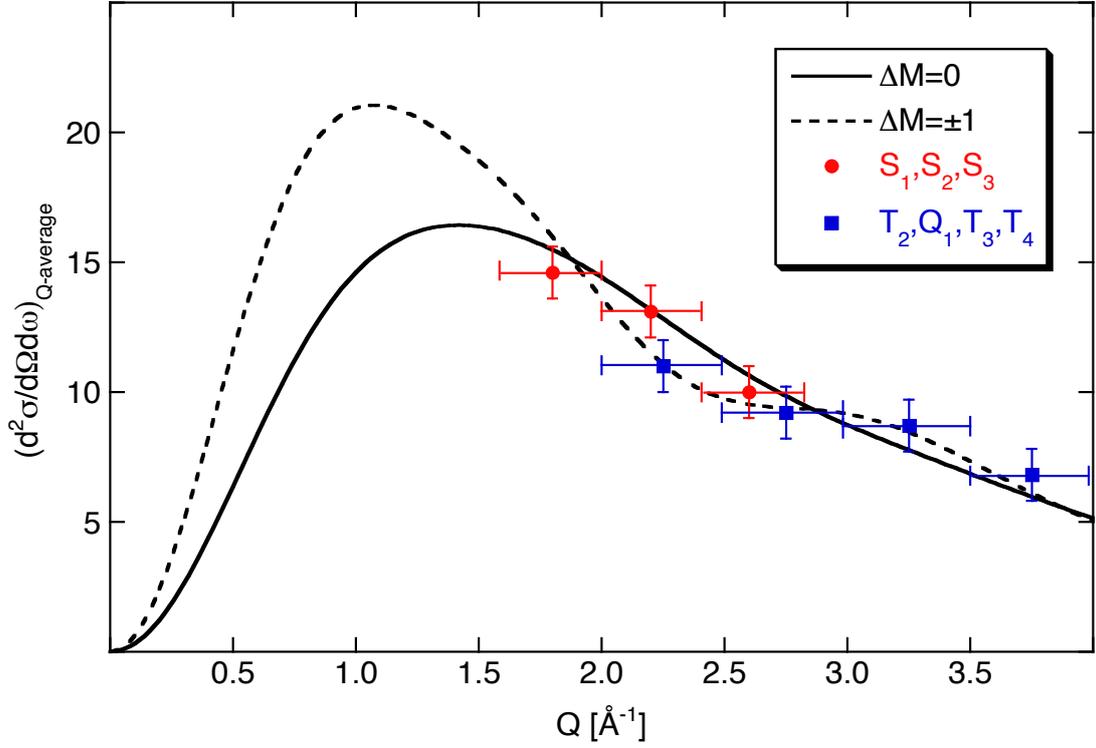

FIG. S1. (Color online) Q-dependence of the neutron cross-section for $Cu^{2+}$ hexamer transitions. The lines denote the calculated Q-averaged cross sections defined by Eq. (2), where all the transition matrix elements $T_{\lambda\lambda'}$ are set equal to 1. The circles and squares correspond to the scaled, integrated intensities of the transitions $S_1$, $S_2$, $S_3$ and $T_2$, $Q_1$, $T_3$, $T_4$, respectively, observed for $K_2Cu_3O(SO_4)_3$ at T=1.5 K.